# Computational fluid dynamic simulation of hull reservoir wave energy device

**V A U De Alwis[1], A P K De Silva[1], S D G S P Gunawardane[1] and Young-Ho Lee[2]**

[1]Department of Mechanical Engineering, University of Peradeniya, Peradeniya, Sri Lanka, sdgspg@eng.pdn.ac.lk

[2]Department of Mechanical Engineering, Korea Maritime and Ocean University, Busan, South Korea, lyh@kmou.ac.kr

**Abstract**. This paper presents a Computational Fluid Dynamics (CFD) analysis of a wave energy device called the Hull Reservoir Wave Energy Converter (HRWEC). The device consists of a floating hull and a flap connected to the shaft of power take-off system (PTO), which is integral to the hull structure. It is unique due to its ability to convert wave energy by utilizing the pitch motion of the hull and rotating flap due to the internal water movement in the hull. Due to the complexity of the internal fluid dynamics, a CFD-based analysis was considered most appropriate. The CFD investigation of the dynamics of the device was done under regular wave conditions by using the ANSYS-AQWA and ANSYS FLUENT. Relative pitch angle variation, the hydrodynamic coefficients, which determine the degree of power extraction, were obtained from simulated results. A simulation was designed exhibiting complete system dynamics for different configurations varying on internal water height. Excellent convergence was observed, and an optimum configuration was identified. It is expected to validate the simulation results through experiments in the foreseeable future.

## 1. Introduction

Ocean waves are a renewable energy source for generating electricity using clean methods. It contains kinetic energy through water motion and potential energy due to the elevation change of water as the waves move. Use of wave energy in power generation has been a subject area open to discussion and research for the past two centuries although substantial progress has not yet been achieved [1]. One of the main reasons for this is most of the novel devices are made to suit the closest sea conditions and when these are tested elsewhere a noteworthy performance was not to be seen. This paper discusses implemented method of CFD analysis of hull reservoir flap type wave energy converter (HRWEC) [2] a hybrid concept of Pendulor wave energy device and Pendulum wave energy converter (PeWEC) [2,3]. These types of novel devices have been the topic for research during recent times and some of the emerging concepts are discussed in following references [4,5,6].

The HRWEC comprises of a floating hull that interacts with the ocean waves, plus the device itself consists of a fluid internally interacting with the walls of the device (see figure 1). The concept of using an oscillating water column inside was mainly inspired from the Pendulor wave energy device which makes use of standing waves and a caisson [7]. An oscillating water column is mainly used to get a variable hydrodynamic effect with changes in frequency. If successful, the device would be able to harness energy for a wider range of frequencies, which in turn would lead to be useful in a wide range of sea conditions. To observe this, this paper focusses on the regular waves because that is the go-to method when it comes to the initial development stages of these devices [8,9].

However, due to the multiple modes of interaction, it is inevitably hard to analyze such a system with many analysis methods. Figure 1 and 2 show the modes representation and artistic impression (interior of the device once outer cover is removed) [2,3] of the HRWEC respectively. However, numerical analysis using CFD codes have been extremely popular due to its ability to model any complicated model and also because results can be extracted with high convergence showing excellent accuracy [10]. With regard to these reasons, this study was centered on identifying a CFD method which would cover the essential dynamics of the system.

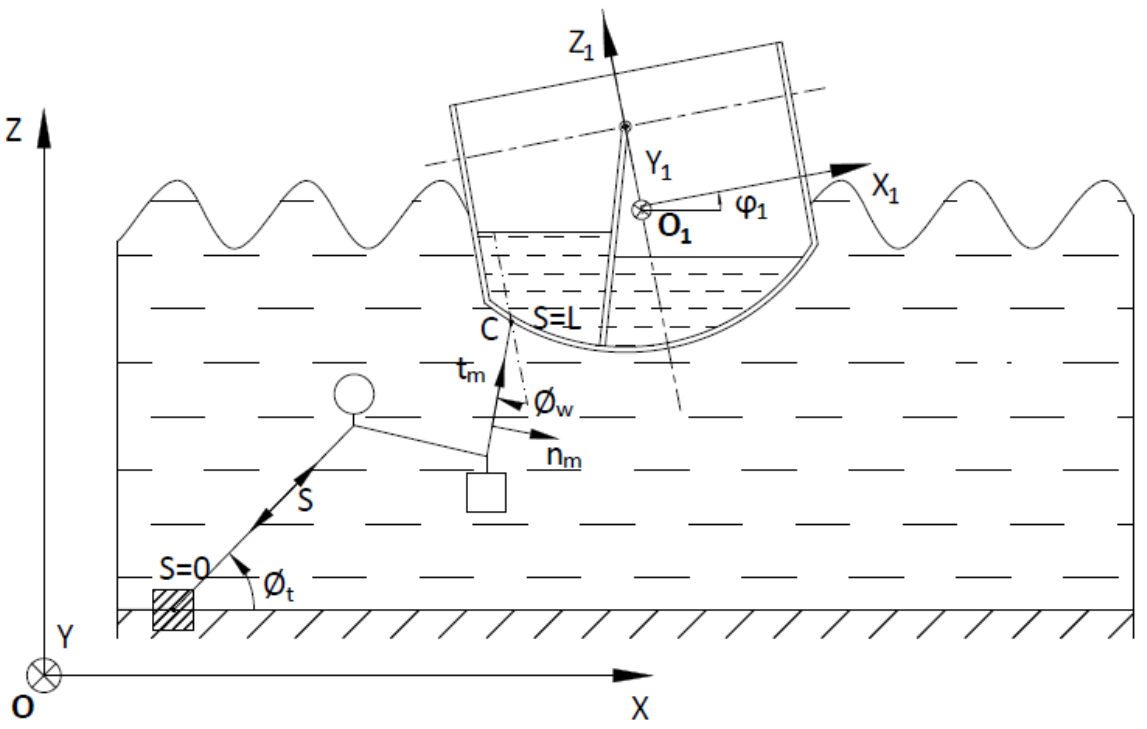

**Figure 1.** HRWEC model- Modes Representation.

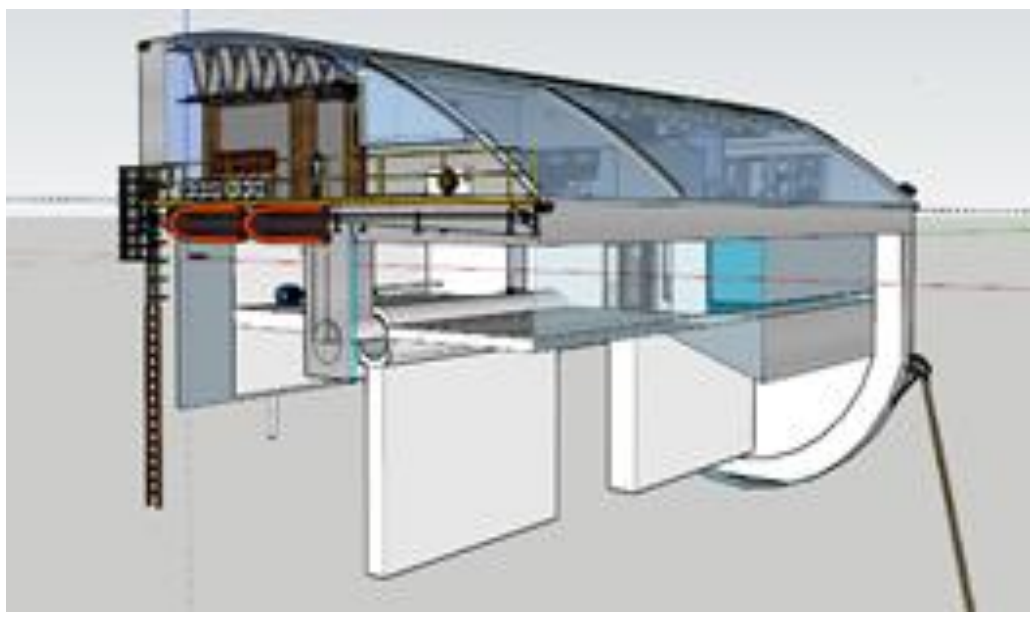

**Figure 2.** HRWEC model Artistic impression.

Due to the symmetric shape of the hull an assumption can be made that the whole motion happens along the surge, heave and pitch directions and hence can be considered to lie on a 2D plane which would in turn neglect sway, yaw and roll movements. Simultaneously the fluid inside the hull controls the relative motion between flap and the hull which is used to generate energy through linear power take off (PTO) damping through possibly a hydraulic power transmission system [11]. Furthermore, different types of forces get generated in the hinge between flap and the hull such as Coriolis, gravitational and reaction forces which lead to its oscillation.

The dependence of internal water inside the hull was studied by using four arbitrary cases for the ease of analysis purposes as shown in table 1 and the parameters described in the table are shown in figure 3. Here, the case 1 is also referred to as the reference case as this case was directly derived from the PeWEC itself [3]. A theoretical model has already been designed for case 1 HRWEC. Due to the availability of a theoretical model which is directly derived from PeWEC case 1 will be referred to as the reference case in this paper discussion.

**Table 1.** Added water level of the hull reservoir wave energy converter (HRWEC).

| Case | Added water volume / (m$^3$) | Internal water height –$X_0$ / (m$^3$) | Submerged height –$Y_0$ / (m$^3$) | Floating height –$Z_0$ / (m$^3$) |
|---|---|---|---|---|
| 1 | 0 | 0 | 0.82 | 1.18 |
| 2 | 2.5 | 0.63 | 1.24 | 0.76 |
| 3 | 3.0 | 0.74 | 1.33 | 0.67 |
| 4 | 3.5 | 0.83 | 1.42 | 0.58 |

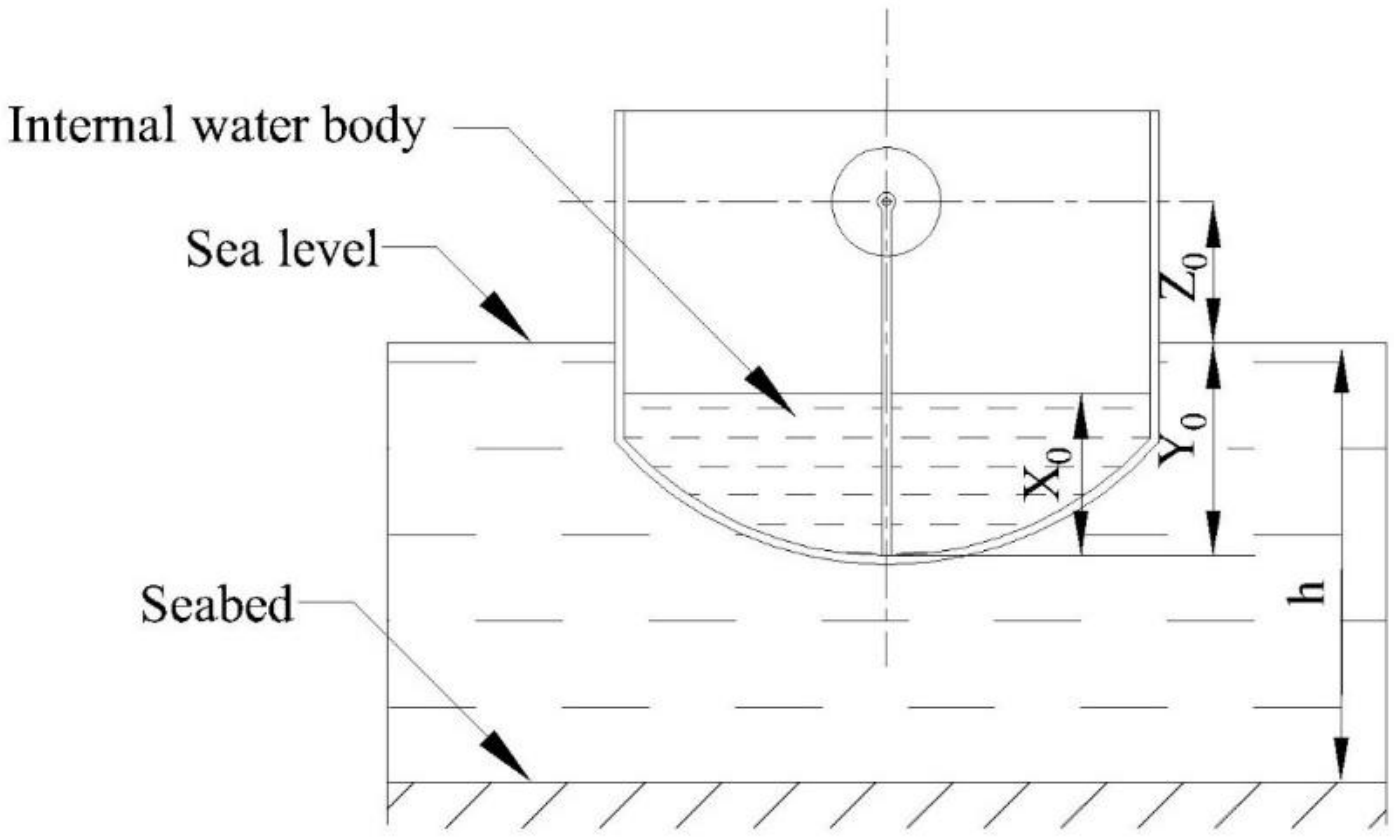

**Figure 3.** Representation of parameters involved in cases to be analysed.

Out of all the commercially available software designed for CFD analysis, Fluent and AQWA, which ANSYS owns, were used for analysis purposes [12,13]. Fluent is a software that is well known for its solver being able to analyze fluid domains to produce high convergence results. Not surprisingly, Fluent could be used to completely model the system, including the wave environment and the motion of the internal fluid. However, such kind of modelling is practically out of hand due to several reasons.

1. Although Fluent is generally designed for all types of fluid analysis, it hasn't incorporated an analysis scheme unique to sea wave analysis.
2. The wave environment would also have to be modelled separately if Fluent is used but still modelling of open sea conditions would be difficult.
3. Modelling the complete system in Fluent would result in a mesh having a large number of unnecessary elements making the simulation complicated and consequently would require high computational power and time.

Due to these reasons, a newer approach was used to break down the complete simulation into two subsections. And for this purpose, an additional software package was used called ANSYS AQWA. The role of each software can be summarized as given below.

1. ANSYS AQWA – This part analyzes the interaction of the device with the wave. (Hydrodynamic diffraction with time response)
2. ANSYS Fluent – This section analyzes the motion of the fluid inside the device. (CFD analysis)

This paper proposes a coupled analysis where AQWA analyzes the hydrodynamic properties when interacting with the wave. The resulting motion is fed into Fluent to analyze the motion of the fluid internally. Although this is a coupled analysis, one of the drawbacks would be the lack of ability to solve both processes simultaneously, which comes at the expense of less costly and faster simulation times. However, this approach has been proved to be a reasonable assumption in various previous literature [14].

## 2. Implemented method and Simulation
The following diagram briefly emphasizes the approach discussed.

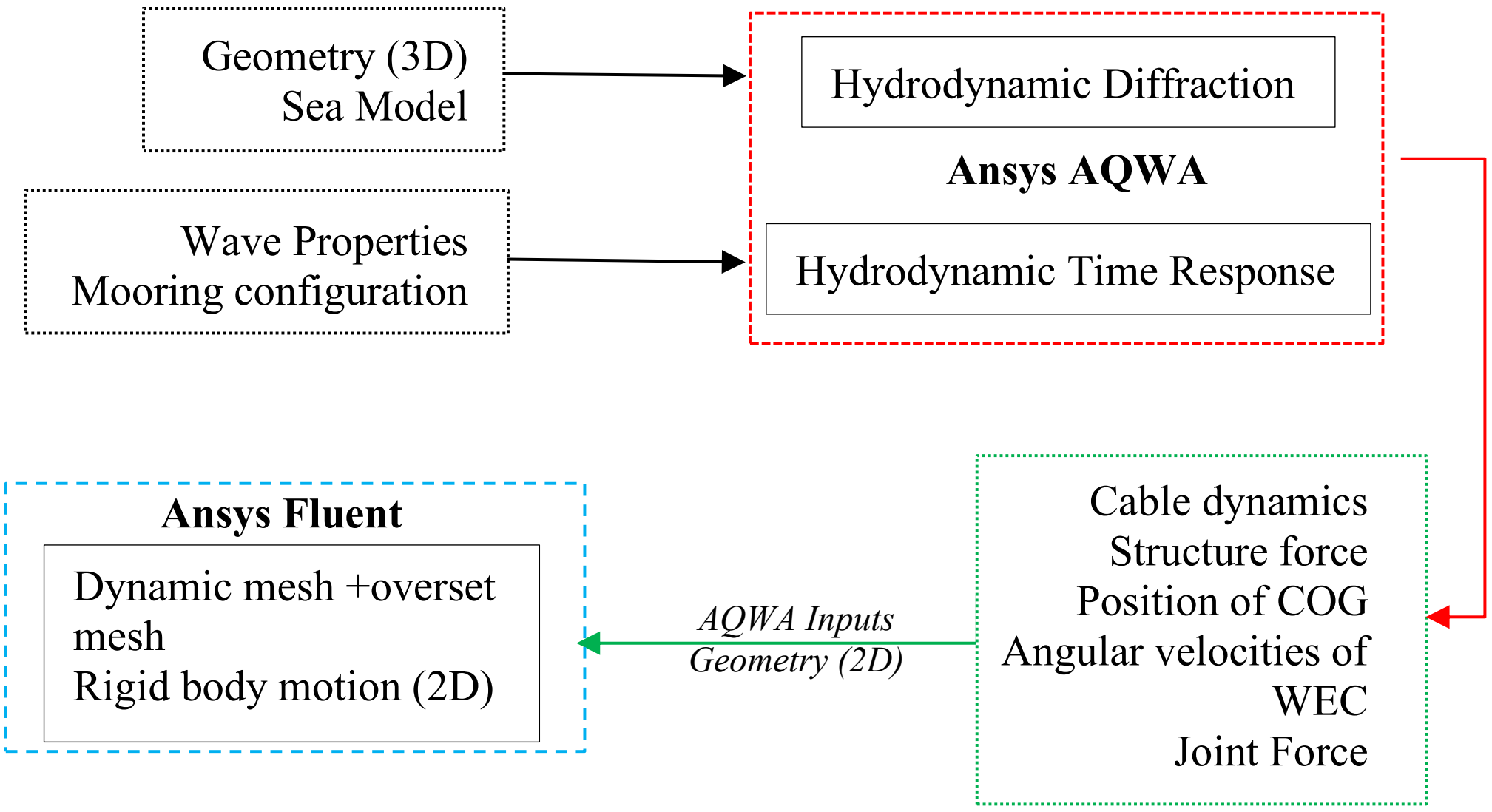

**Figure 4.** Schematic diagram of implemented method for simulation of HRWEC.

Figure 4 illustrates AQWA and fluent approach for hydrodynamic and CFD analysis of HRWEC. The main objective of AQWA analysis is to capture the motion of the flap and the hull of WEC within a given wave condition. Table 2 & 3 indicate the inputs for ANSYS AQWA model which were taken to exactly match them to the parameters of the PeWEC [2,3,15] and the parameters involved in the two tables are shown in figure 5. Features and dimensions of the HRWEC are given in table 2. Regular wave properties, mooring configuration and sea model details are shown in table 3. The hydrodynamic response data for Flap (time & pitch angular velocity) and for Hull (time & heave, surge linear velocities, pitch angular velocity) are exported to a .csv File for cases in table 1. These data were taken with respect to Global coordinate system and used as inputs for the analysis in ANSYS Fluent. The main aim of using Fluent in this analysis is to analyze the internal water sloshing properties of fluid domain and to see the behavior of phase contours.

**Table 2.** Features and dimensions of the hull reservoir wave energy converter (HRWEC).

| Description | Value |
|---|---|
| **Hull** | |
| Length (x-direction) | 3 m |
| Width (y-direction) | 2 m |
| Height (z-direction) | 2 m |
| Radius | 1.5 m |
| Mass of the hull | 3176 kg |
| Roll axis moment of inertia | 1499 kg/m2 |
| Pitch axis moment of inertia | 2168 kg/m2 |
| Yaw axis moment of inertia | 2761 kg/m2 |
| **Flap** | |
| Mass | 620 kg |
| Moment of inertia with respect to the hinge point | 360 kg/m$^2$ |
| Length | 1.95 m |
| Thickness | 0.05 m |
| Width | 1.90 m |
| Clearance between the hull and the flap | 2 cm |

**Table 3.** Sea model, mooring configuration and Regular wave properties.

| Description | Value |
|---|---|
| Wave depth | 3.5 m |
| Width (y-direction) | 8 m |
| Length (x-direction) | 10 m |
| Net gravity force | 100 N |
| Net buoyancy force | 250 N |
| Length of line 1 (from seabed) | 1.8 m |
| Length of line 2 | 0.5 m |
| Length of line 3 | 2.0 m |
| Wave period | 2.2 s |
| Wave amplitude | 0.15 m |

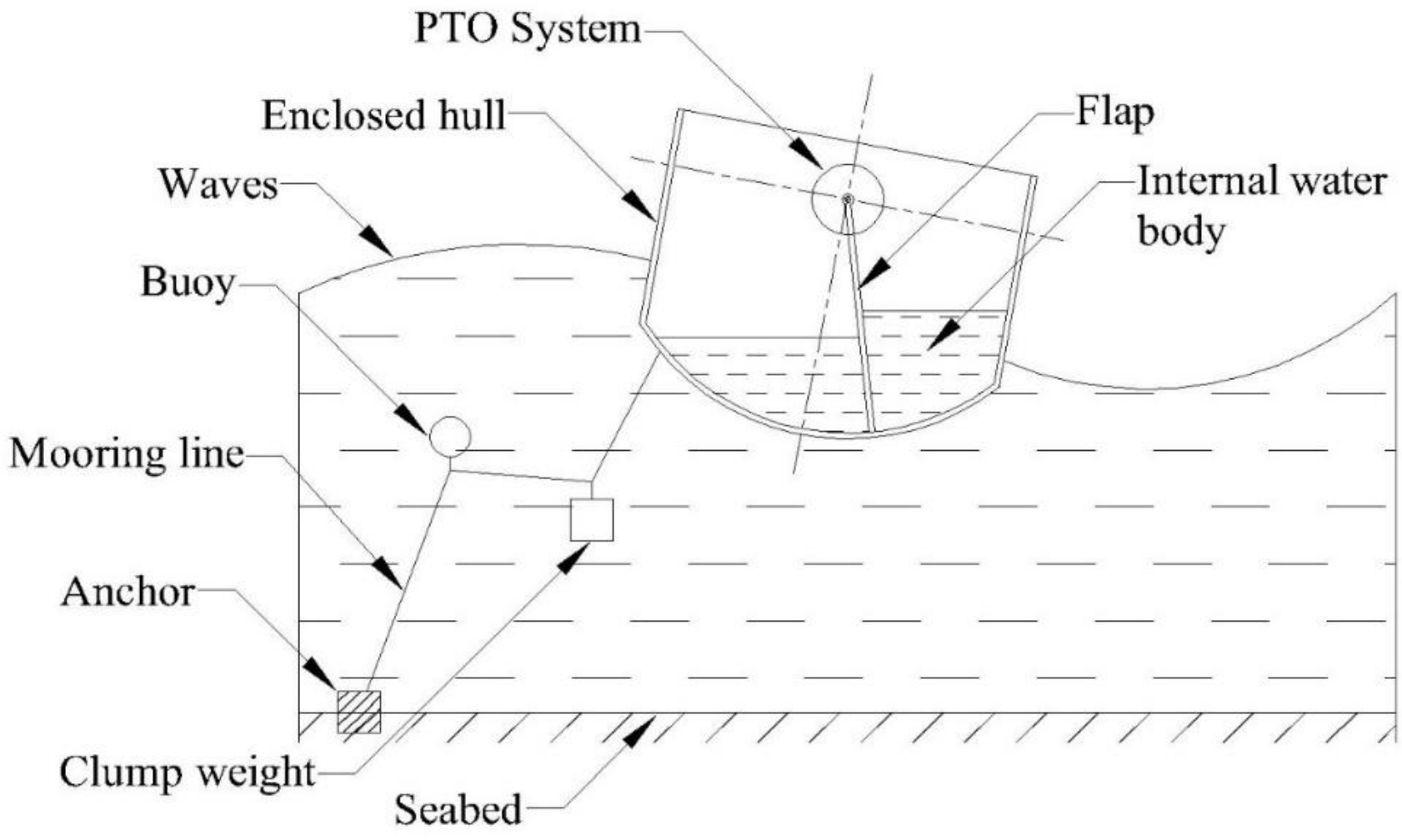

**Figure 5.** Complete representation of the simulation setup.

## 3. Simulation model

*3.1 AQWA model*

HRWEC is modeled as a rigid body with six DOF system for motion. It is presumed that the wave transmits along the surge axis of HRWEC. Due to the symmetricity of the hull, it is considered that roll motion, sway motion and yaw motion are zero. Therefore, the hull has modes only in surge, heave, and pitch directions and the wave action is excited on the hull in these directions. AQWA is used to perform the hydrodynamic analysis of HRWEC. Mooring configuration, Regular wave pattern and sea model are considered as the inputs to the AQWA and heave motion, surge motion, pitching speed and cable dynamics of the HRWEC were obtained. AQWA setup is shown in figure 6. It was clear after the AQWA simulation that the model behaves as expected where roll, sway and yaw motions are negligible.

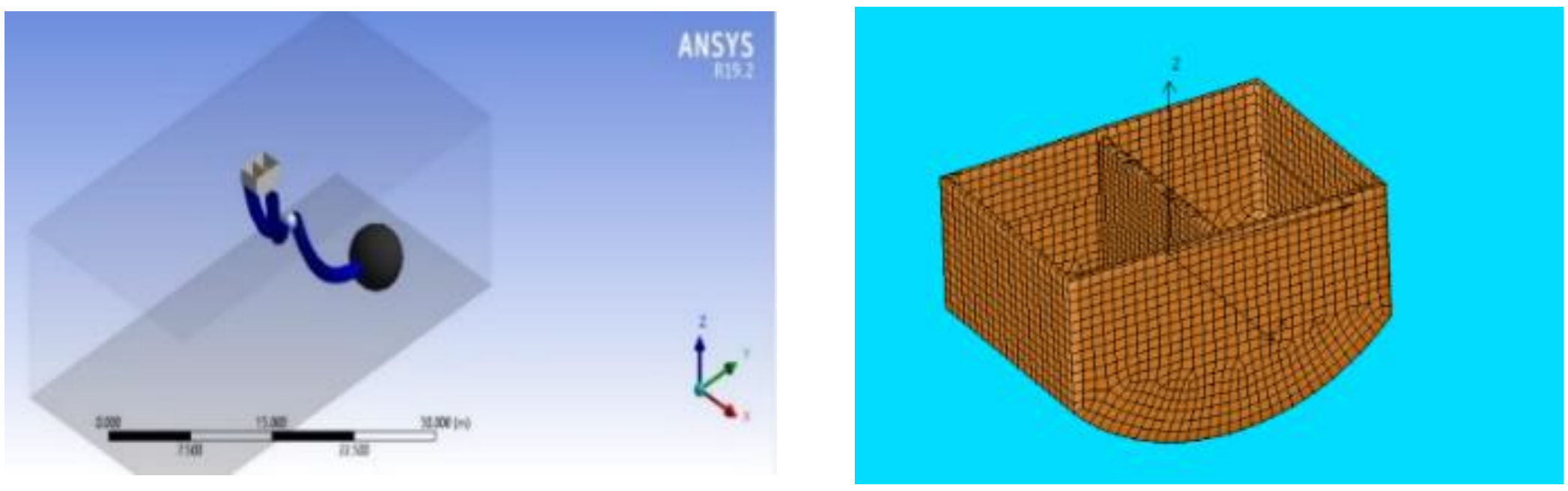

**Figure 6.** Ansys AQWA Setup.

*3.2 Fluent model*

For regular waves, when analyzed in AQWA, only pitch surge and heave motions of the hull could be seen. Therefore, the motion of the device can be considered to reside in a two-dimensional plane. Also, the wave energy device has a uniform cross section along its rotating axis. Therefore, a two-dimensional fluid domain can be used to carry out the fluid analysis, which would save a substantial amount of simulation time.

**Mesh.** As shown in figure 7 and 8, two bodies are present in the fluid domain, and both of them need to be meshed. Here, two separate meshes were created for both the flap and the hull, and the flap was modelled as a wall inside the fluid domain. Once the meshes for the two bodies are constructed, they transform into a single mesh with a moving wall when the two mesh interfaces are intersected. This phenomenon is called overset meshing [12]. The separate meshes applied for the two bodies can be shown as in figure 7, and once the overset interfaces are intersected, the whole fluid domain converges to a single mesh, as shown in figure 8.

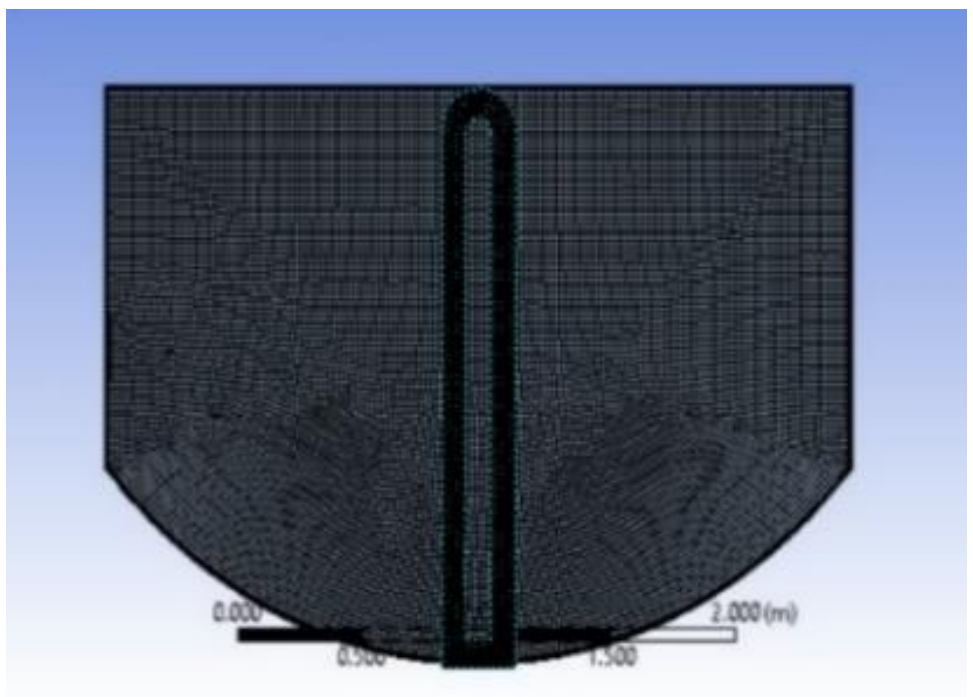

**Figure 7.** Overset mesh.

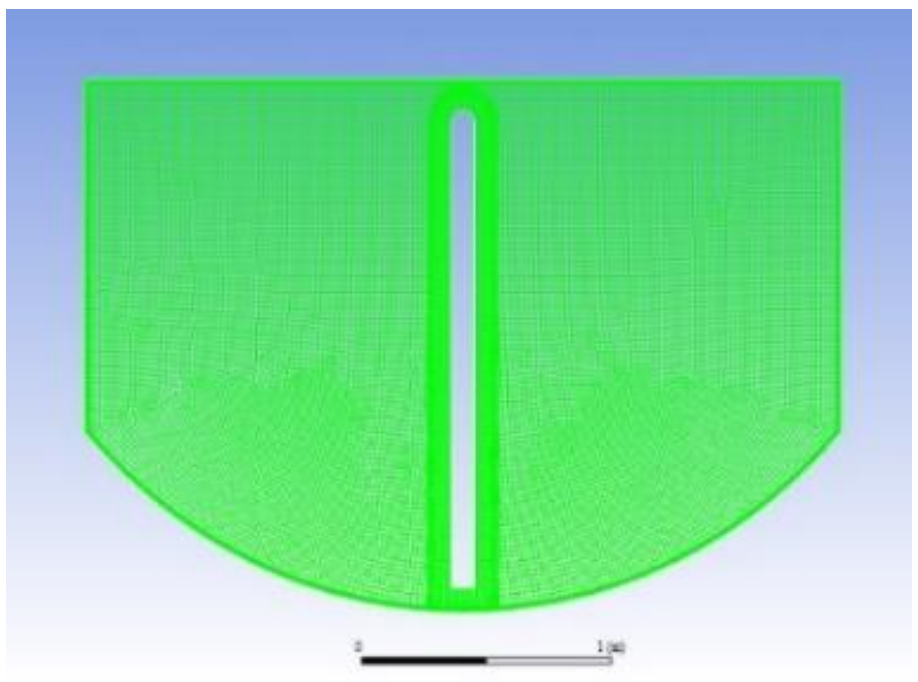

**Figure 8.** Overset fluid domain.

**Setup details.** In the analysis, as the fluid domain was separated into two regions, namely as air and water, a Multiphase model needed to be specified. Here a Volume of fluid model was considered as the most appropriate [12]. Furthermore K-epsilon model was used as a viscous model to analyze the turbulence occurring inside the device.

## 4. Theory

ANSYS-AQWA simulator is based on the potential flow and diffraction theory. This simulator assumes an incompressible, inviscid, and irrotational flow [13].

*4.1 Instantaneous power absorbed by the device,* $P_{d,PTO}$

The energy capture of the device happens at the power take off unit (PTO) attached at the hinged of the flap. In this analysis, it is assumed that equivalent viscous damper is used as the PTO. Therefore, the instantaneous power absorbed by the device [16],$P_{d,PTO}$ at the hinge joint and can be calculated from dot product of instantaneous torque $\vec{\tau}(t)$ and relative angular velocity $\vec{\omega}(t)$ of flap and hull as shown in equation (1).

$$P_{d,PTO} = \vec{\tau}(t) \cdot \vec{\omega}(t) \qquad (1)$$

The Root-Mean-Square (RMS) power (mean power), $P_{RMS,PTO}$ at the hinge joint can be calculated from equation (2): where $N$ is number of samples in the time series.

$$P_{RMS,PTO} = \sqrt{\sum_{i=1}^{N} P_{d,PTO}} \qquad (2)$$

*4.2 Capture factor,* $C_f$

Generally, the capture factor $C_f$ is used to represent the efficiency of a wave energy converter [17]. It is defined as shown in equation (3), where D is the effective incoming wave width.

$$C_f = \frac{P_{RMS,PTO}}{P_i D} \qquad (3)$$

## 5. Results and Analysis

*5.1 Fluent results*

The analysis in Fluent was carried out to mainly interpret the internal sloshing action of fluid inside. In this section, more properties related to the fluid inside are analysed in depth with the aid of hydrodynamic parameters. As seen in the curves initially high drag and moment values are seen. However, as time progresses the device stabilizes showing periodic sinusoidal variation. To compare the performance of all the cases the analysis was further extended similarly and the curves of drag forces and moments of the hinge were plotted on the same axes as shown in figure 9 and 10.

According to the figures, it is evident that the drag forces have reduced slightly while moments have increased with the water volume suggesting that the degree of wetted surfaces have led to this behaviour. Furthermore, these coefficients depend on the gap that is present between the bottom of the flap and hull and a small change in the clearance would complicate the motion to a large extent. Hence an optimum clearance should be kept depending on the case selected based on other criteria.

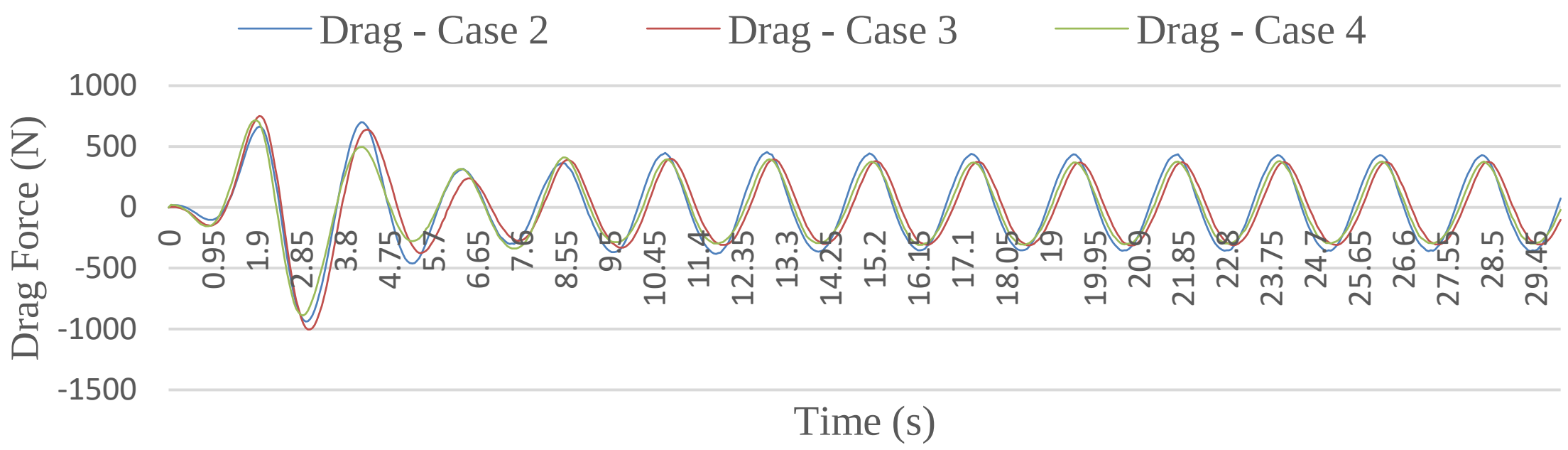

**Figure 9.** Variation of Drag force with time for the drag coefficient generating maximum efficiency of each case.

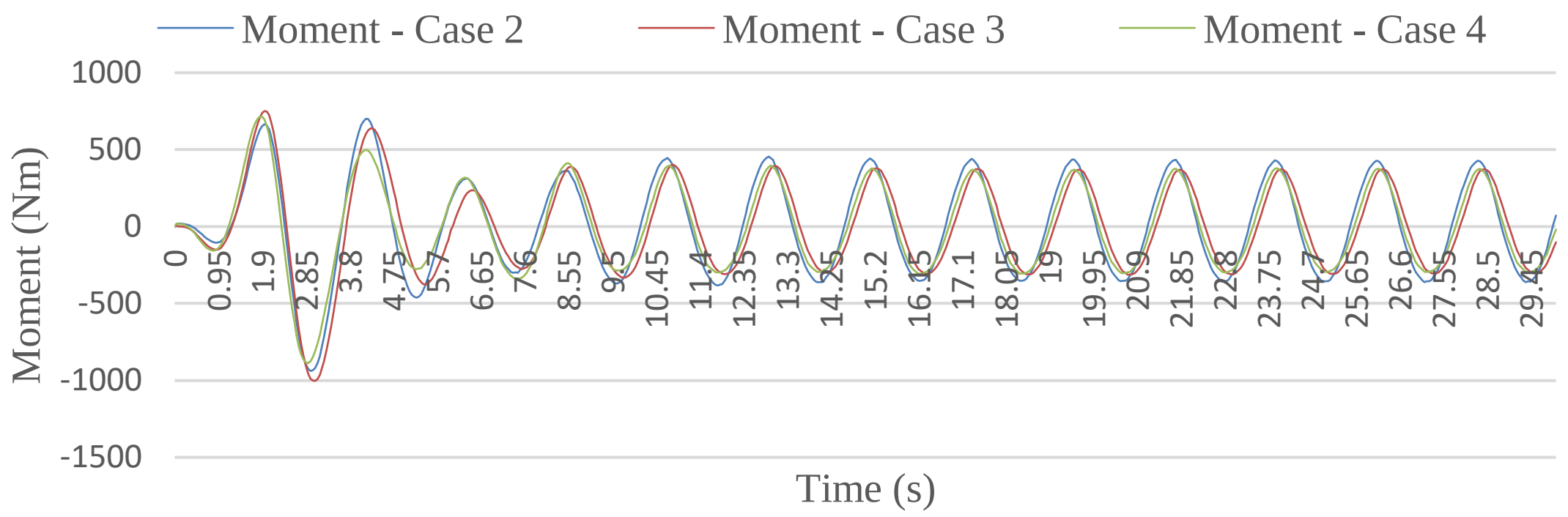

**Figure 10.** Variation of moment with time for the drag coefficient generating maximum efficiency of each case.

Furthermore, when sloshing takes place inside the hull it is obvious that there will be kinetic energy dissipating due to turbulence. For the study of this, contours of turbulent kinetic energy were obtained during the simulation and identical time steps were compared for the turbulent kinetic energy dissipation. In figure 11 turbulent kinetic energy after 12.5 seconds is demonstrated for each case with the phase behaviour corresponding to each of them shown at the same time.

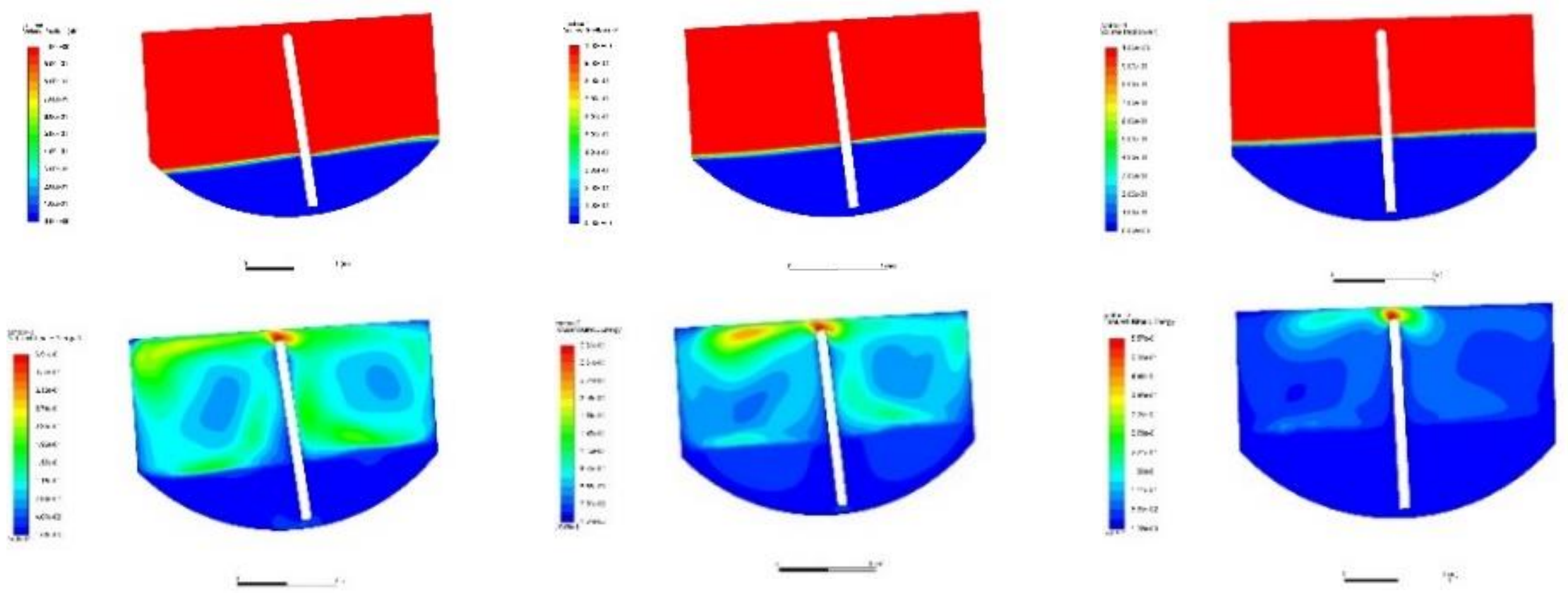

(a) (b) (c)

**Figure 11.** Phase and turbulent intensity contours for Case 1(a), Case 2(b), Case 3(c) at 12.5 seconds simulation time.

Looking at the contours of turbulent kinetic energy it is evident that there is a separation in values between the water and air volumes. Moreover, it is seen that the energy dissipation in the water region has significantly lower values compared to that of air suggesting as water volume increases kinetic energy dissipation decreases. These factors should also be taken into consideration when determining an optimum water volume inside the hull.

*5.2 AQWA results analysis*

Figure 12 shows variation of capture factor for variations of PTO damping coefficient (0-600 N.m.s / rad) for HRWEC reference case and PeWEC [3]. For this result, the HRWEC was simulated under a regular wave with 0.15 m amplitude and 2.2 s wave period which are the same conditions used for the experimental analysis of PeWEC done in the Mediterranean Sea, which is well known for its waves with high frequency and low amplitude [18,19]. By looking at the graph it is evident that the analysis results obtained for the HRWEC are similar to the Capture factors of experimentally performed PeWEC both qualitatively and quantitatively suggesting that the simulation results would yield acceptable results even for the other cases considered in the analysis. Hence, implemented CFD model can assumed as an accurate model.

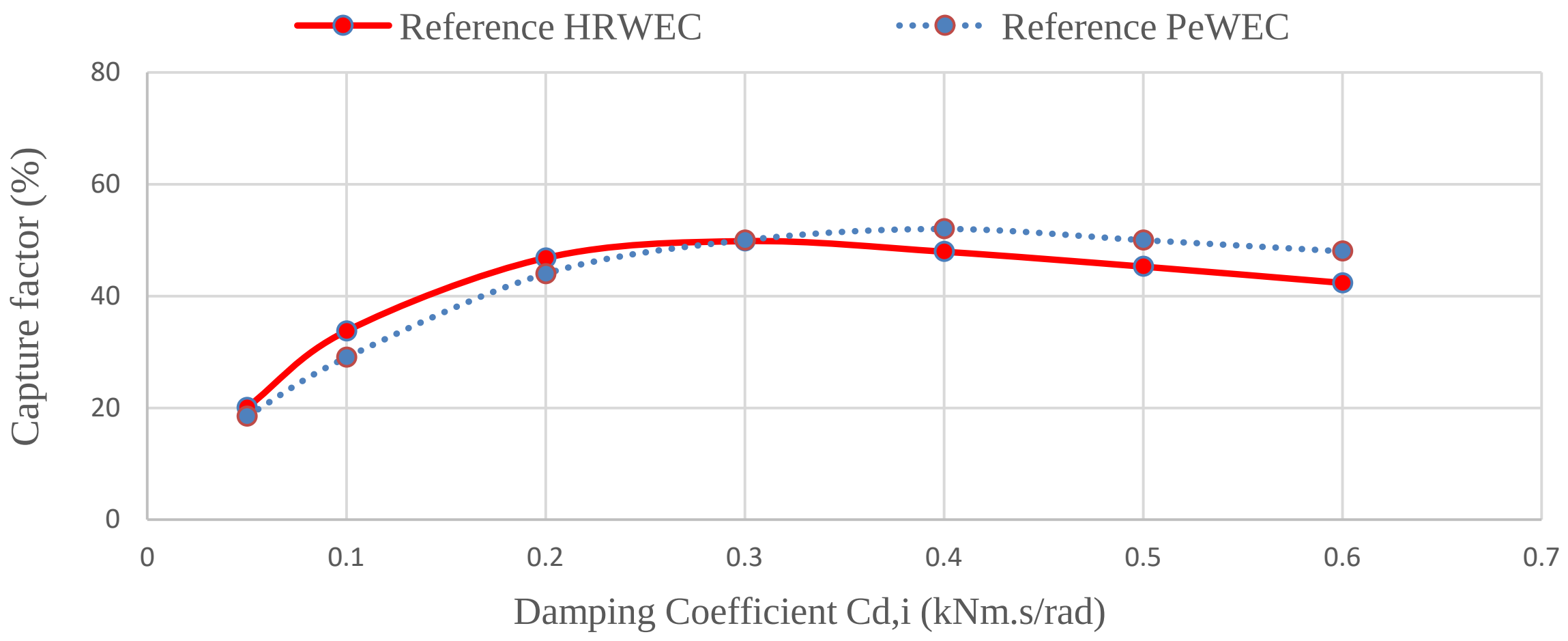

**Figure 12.** Capture factor with various PTO damping coefficient for HRWEC and PeWEC.

This method of analysis was then extended to the cases where water was filled inside the hull according to those respective cases. Figure 13 shows variation of capture factor for HRWEC for variations of PTO damping coefficient (0-30 kN.m. s/ rad) for cases 1,2,3 and 4. The device was simulated under the same wave properties as mentioned. The Root-Mean-Square (RMS) values are used to represent the values of absorbed powers in figure 13.

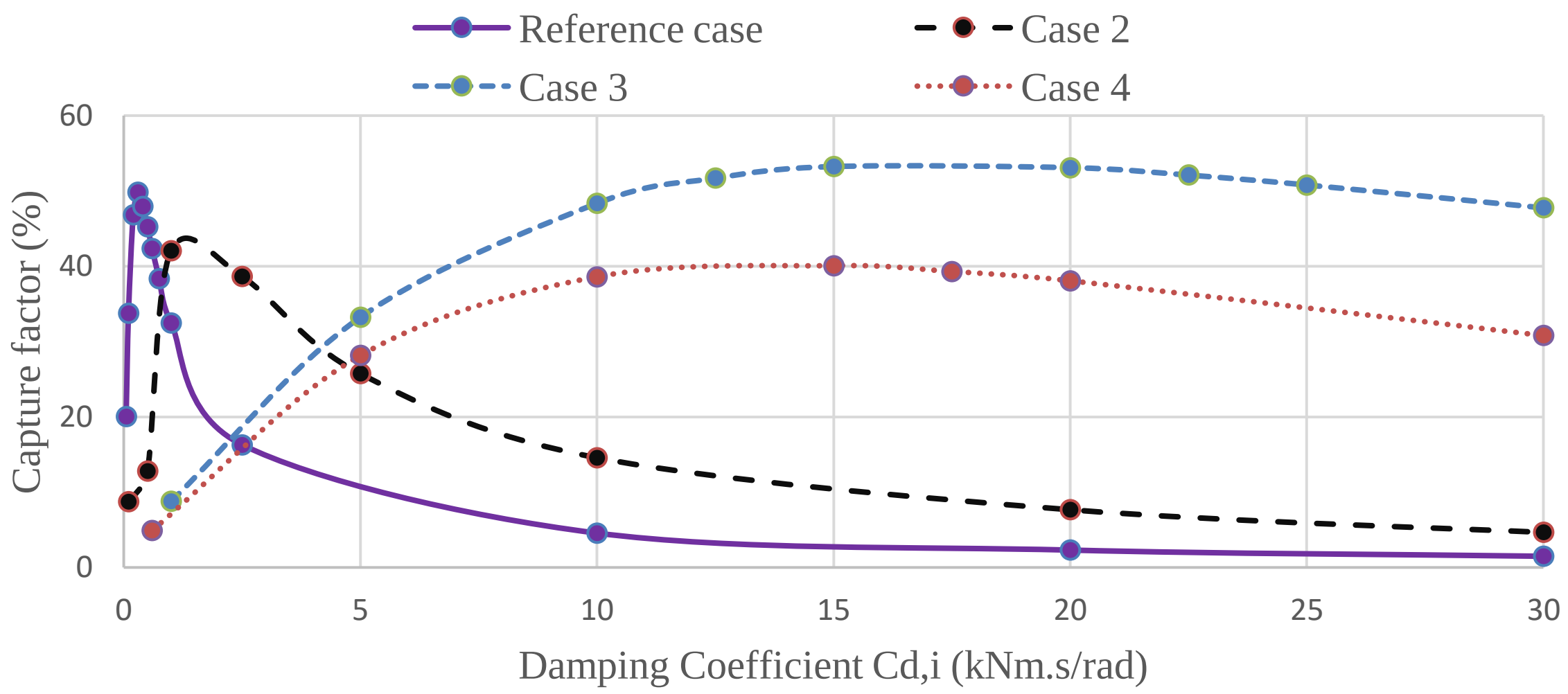

**Figure 13.** Variations of capture factor with various PTO damping coefficient for each case.

The capture factor increases with increasing damping coefficient, and then decreases after attainment of a maximum value. Therefore, optimal damping coefficient for maximizing the capture factor is essential. The reason for this tendency is due to the fact that power captured is the product of damping torque and angular velocity (refer section 4.1). A low damping coefficient will result in high angular velocity but small torque, whereas a high damping coefficient will result in low angular velocity but high torque. Consequently, compromise must be found for the best power capture and capture factor possible.

The maximum capture factors were obtained at 0.3,1, 15 and 15 kN.m.s/ rad and capture factors were 49.8%, 42%,53.2% and 40.1% for reference case, case 2, case 3 and case 4 respectively. According to the figure 13, it is clear that the case 3 has best performance under the regular wave conditions for higher damping coefficients [2] while case 1 has better performance for smaller damping coefficients. These results suggest that the inclusion of a fluid inside the hull has an impact towards its capture factor. Moreover, at lower internal water levels and also at very high internal water levels the device would exhibit a lower performance than its parent case excluding water. Hence it is vital to do further research on what internal fluid levels to be maintained.

## 6 Conclusion

The study was done on a recently introduced wave energy device which is shortly called the Hull Reservoir Wave Energy Converter and its performance was analyzed using CFD tools which provide numerous benefits where analysis under fluid motion is considered. After considering different approaches suitable for analysis, it was decided to use two software namely AQWA and Fluent produced by ANSYS for simulation of wave conditions and internal water dynamics respectively. Results obtained were compared to PeWEC and HRWEC in previous literature to show the validity of the study, and to analyze power generation as carried out in this paper.

Initially it was proved that the method of analysis used would provide reliable results by comparing it to the experimental analysis of PeWEC. Accordingly internal water sloshing was studied in Fluent for three arbitrary cases and it was found that drag force decreases with water volume and the moments showed contrasting behavior. Furthermore, turbulence was studied and it was concluded that inclusion of water would reduce the energy dissipation. Finally, it was found that filling the hull with water would change the efficiency of the device and accordingly it was found that the case 2 produce the best capture factor, even greater than that of the reference case. Hence it was concluded that an optimum water level needs to be identified for the best power generation. However, all these results are expected to be analyzed using experiments in near future.